\documentclass[letterpaper, 10 pt,conference]{ieeeconf}
\IEEEoverridecommandlockouts                              
\usepackage{epsfig}
\usepackage{amsmath}
\usepackage{amssymb}
\usepackage[T1]{fontenc}
\usepackage{tikz}
\usetikzlibrary{shapes,arrows}
\usepackage{pgfplots}
\pgfplotsset{plot coordinates/math parser=false}
\usepackage{algorithmic}
\usepackage{algorithm}
\usepackage{dsfont}
\usepackage{xspace}
\usepackage{graphicx}
\usepackage{subfig}

\def\ie{{\em i.e.,}\xspace}
\def\cf{{\em c.f.,}\xspace}

\definecolor{kthblue}{cmyk}{1,0.55,0,0}
\definecolor{kthred}{cmyk}{0,1,0.65,0.23}
\definecolor{kthgreen}{cmyk}{0.28,0,0.70,0.3}

\newtheorem{remark}{Remark}

\newcommand{\fig}[1]{Figure~\ref{#1}}
\newcommand{\LMI}{\textsc{lmi}\xspace}
\newcommand{\SDP}{\textsc{sdp}\xspace}
\newcommand{\MPC}{\textsc{mpc}\xspace}
\newcommand{\FIR}{\textsc{fir}\xspace}
\newcommand{\PEM}{\textsc{pem}\xspace}
\newcommand{\CVX}{\textsc{cvx}\xspace}
\newcommand{\MIMO}{\textsc{mimo}\xspace}

\title{\LARGE \bf
On Optimal Input Design for Feed-forward Control
}
\author{Per Hägg and Bo Wahlberg
\thanks{This work was partially supported by the European Research Council under the European Community's
  Seventh Framework Programme (FP7/2007-2013) / ERC Grant Agreement No. 267381, the Swedish Research Council and the Linnaeus Center
ACCESS at KTH}
\thanks{The authors are with the Automatic Control Lab and ACCESS, School of Electrical Engineering, KTH, The Royal Institute of Technology, SE-100 44 Stockholm, Sweden. \newline
E-mail: { \{pehagg, bo\}@kth.se }
}}

\begin{document}

\maketitle
\thispagestyle{empty}
\pagestyle{empty}

\begin{abstract}
This paper considers optimal input design when the intended use of the identified model is to construct a feed-forward controller based on measurable disturbances. The objective is to find a minimum power excitation signal to be used in system identification experiment, such that the corresponding model-based feed-forward controller guarantees, with a given probability, that the variance of the output signal is within given specifications.
To start with, some low order model problems are analytically solved and fundamental properties of the optimal input signal solution are presented.  The optimal input signal contains feed-forward control and depends of the noise model and transfer function of the system in a specific way.  Next, we show how to apply the partial correlation approach to closed loop optimal experiment design to the general feed-forward problem. A framework for optimal input signal design for feed-forward control is presented and  numerically evaluated on a temperature control problem.
\end{abstract}

\section{Introduction}
\PARstart{S}{ystem} identification is about the estimation and validation of mathematical models of dynamical systems from experimental data. It is well recognized that the input signal used to excite the
system during the experiment significantly affects the accuracy of the identified model. A correctly chosen excitation signal can immensely improve the quality of the resulting model while a poorly chosen signal can result in a useless model. This motivates input or experiment design in system identification.

The quality of a model depends on the intended use of the model. A good experiment should highlight the important properties for the intended application.  This is the main idea of \emph{identification for control} \cite{Gevers:2005,Gevers:1986}, \emph{least costly identification} \cite{Bombois:2006} and \emph{applications oriented input design} \cite{Hjalmarsson:2009}.

Much work has been focused on input design when the model is to be used in a control application, see for example \cite{Wahlberg:2010} or \cite{Larsson:2011} for application to \MPC. However, many industrial control systems do not only utilize feedback but also feed-forward control. The idea is to measure a disturbance, predict its impact on the plant and then compensate for it with the input signal. Compensating for measurable disturbances by feed-forward control can improve the performance considerably compared to when only feedback control is used. Measurable disturbances could for example be the outside temperature when controlling the indoor temperature in a house.

The aim of this paper is to study optimal input design when the intended use of the model is for feed-forward control. We will use the application oriented input design framework presented in \cite{Hjalmarsson:2009}. To illustrate how system properties affect the optimal input signal we start by analyzing a problem which can be solved analytically. We then present a framework for input design for more general systems and show how to formulate them as Semi Definite Programs (\SDP) that can be solved efficiently using numerical methods.

The outline of the paper is as follows. In Section~\ref{sec:problem} we define the input design problem and give some preliminary results from application oriented input design. Section~\ref{sec:FIR} analyzes first order \FIR subsystems and make some observations. A more general framework for input design for feed-forward control is given in Section~\ref{sec:general}. The framework is then applied to a simulation example in Section~\ref{sec:example} while Section~\ref{sec:conclusions} concludes the paper.

\section{Problem Formulation}\label{sec:problem}
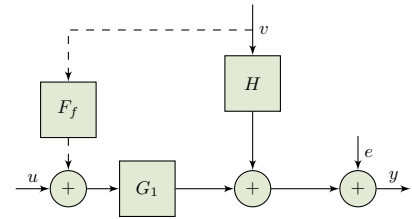
\begin{figure}[h]
\centering
{\pgfplotsset{width=0.3\columnwidth,height=0.1\columnwidth,compat=newest,plot coordinates/math parser=false}\tikzstyle{block} = [draw, fill=kthgreen!20, rectangle,
    minimum height=3em, minimum width=3em,align=center]
\tikzstyle{sum} = [draw, fill=kthgreen!20, circle, node distance=1cm]
\tikzstyle{input} = [coordinate]
\tikzstyle{output} = [coordinate]
\tikzstyle{pinstyle} = [pin edge={to-,thin,black}]

\begin{tikzpicture}[auto, node distance=1cm,>=latex',transform shape,scale=0.7]

    \node [input] (inputu) {};
    \node [sum,right of=inputu,node distance=1cm](sum1) {$+$};
    \node [block,right of=sum1,node distance=1.5cm] (G1) {$G_1$};
    \node [sum,right of=G1,node distance=2cm] (sum) {$+$};
    \node [block,above of=sum,node distance=2cm] (H) {$H$};
    \node [input,above of=H,node distance=1.5cm] (inputv) {};
    \node [sum,right of=sum,node distance=2cm] (sum2) {$+$};
    \node [input,above of=sum2,node distance=1cm] (inpute) {};
    \node [output,right of=sum2,node distance=1cm] (output) {};
    
    \node [block,above of=sum1,node distance=1.5cm] (Ff) {$F_f$};

    \draw [draw,->] (inputu) -- node[]{$u$} (sum1);
    \draw [draw,->] (sum1) -- (G1);
    \draw [draw,->] (G1) -- (sum);
    \draw [draw,->] (H) -- (sum);    
    \draw [draw,->] (inputv) -- node[name=v]{$v$} (H);
    \draw [draw,->] (sum) -- (sum2);
    \draw [draw,->] (sum2) -- node[]{$y$} (output);    
    \draw [draw,->] (inpute) -- node[]{$e$} (sum2);
    
    \draw [draw,dashed,->] (Ff) -- (sum1);
    \draw [draw,dashed,->] (v) -| (Ff);

\end{tikzpicture}}
\caption{The considered system. The identification setup without feed-forward filter ($F_f$) and application setup with feed-forward filter.}
\label{fig:feed-forward}
\end{figure}

Consider the system in \fig{fig:feed-forward}.
The dynamics of the system can be written as
\begin{equation*}
    y_t = G_0(q)u_t + H_0(q)v_t + e_t,
\end{equation*}
where $q^{-1}$ is the delay operator, \ie $q^{-1}u_t = u_{t-1}$.
Here $G_0(q)$ and $H_0(q)$ are the plant transfer functions. The input to the system is $u_t$ while $y_t$ is the measured output disturbed by a zero mean white noise signal $e_t$ with variance $\lambda_e$. The process $v_t$ corresponds to a measurable disturbance to the system. We assume that $v_t$ is a stationary stochastic process with known spectral properties, \ie $v_t$ can be written as $v_t = M(q) s_t$ where $s_t$ is a zero mean Gaussian process with variance $\lambda_s$ and the stable and inversely stable filter $M(q)$ is known. The spectrum of $v_t$ equals $\Phi_v(\omega) = |M(e^{j\omega})|^2 \lambda_s$.

The objective is to design a feed-forward controller $F_f$ in \fig{fig:feed-forward} to suppress the effect of the measurable disturbance in $y_t$. However, the dynamics of the subsystems $H$ and $G$ are assumed unknown and need to be identified. Hence we want to identify the system dynamics and use the identified model to design a feed-forward controller to counteract the influence of the measured disturbance $v_t$ on the output $y_t$. In particular, we will study to design the input $u_t$ during the experiment such that we with high probability can guarantee that the designed feed-forward controller satisfy an accuracy constraint.

We parameterize the submodels as
\begin{equation*}
    G(q,\theta),\, H(q,\theta),
\end{equation*}
where $\theta \in \mathds{R}^n$ is the model parameter vector and we assume that there is a vector $\theta_0$ such that
\begin{equation*}
    \begin{aligned}
        G(q,\theta_0) = G_0(q), \,H(q,\theta_0) = H_0(q),
    \end{aligned}
\end{equation*}
that is, the true system can be described by the model.

We denote the model parameter vector estimated using a Prediction Error Method (\PEM), \cite{Ljung:1999}, from $N$ data points of the inputs and outputs, $\left\{u_t,v_t,y_t,t=1,\ldots,N \right\}$  as $\hat \theta_N$.
Under some mild assumptions the asymptotic (in $N$) covariance matrix of the estimated parameters
\begin{equation*}
    P = \lim_{N \rightarrow \infty} NE\left\{(\hat \theta_N-\theta_0)(\hat \theta_N-\theta_0)^T\right\},
\end{equation*}
can be written as
\begin{equation}
    P^{-1} = \frac{1}{2\pi \lambda_e}\int_{-\pi}^{\pi} \mathcal{F}_0(e^{j\omega}) \Phi_{\chi_0}(\omega)\mathcal{F}_0^*(e^{j\omega}) d\omega,
    \label{eqn:pinv}
\end{equation}
where the joint spectrum
\begin{equation*}
    \Phi_{\chi_0}(\omega) = \begin{bmatrix}
        \Phi_u(\omega)  & \Phi_{us}(\omega) \\
        \Phi^*_{us}(\omega) & \lambda_s(\omega) \\
    \end{bmatrix},
\end{equation*}
is the spectrum of $\chi_0 = \begin{bmatrix}u_t & s_t\end{bmatrix}^T$ and
\begin{equation}
    \mathcal{F}_0(q) = \left. \begin{bmatrix} \frac{\partial G(q,\theta)}{\partial\theta} & M(q)^*\frac{\partial H(q,\theta)}{\partial\theta}  \end{bmatrix} \right|_{\theta = \theta_0},
    \label{eqn:predict_grad}
\end{equation}
see \cite{Ljung:1999} for details.
\begin{remark}
Since we can measure $v_t$, we can correlate the input $u_t$ with $s_t$ (since we know $M(q)$) and hence $\Phi_{us}$ could be non-zero.
\end{remark}

Further we know that, asymptotically, the identified parameter vector $\hat \theta_N$ lies within the set
\begin{equation*}
    \mathcal{E}_{SI} = \left\{ \theta : \frac{1}{2}\left[\theta-\theta_0\right]^T P^{-1} \left[\theta-\theta_0\right] \leq \frac{\kappa}{N} \right\}
\end{equation*}
with probability $\alpha$. The constant $\kappa$ can be determined from the $\chi^2(n)$ distribution as a function of the probability $\alpha$ and $n$, the number of estimated parameters.

\subsection{Application oriented input design}
As mentioned earlier we would like to identify the parameters of the system and from this design a feed-forward controller. The feed-forward controller will be denoted $F_f(q,\theta)$.

Using the framework in \cite{Hjalmarsson:2009}, we let $V_{app}(\theta)$ denote an application cost that measures the degradation in performance due to model errors. Here the application cost will represent the output variance from the measured disturbance $v_t$ when designing the feed-forward controller with the identified parameters instead of the true parameters, \ie all acceptable models satisfy
\begin{equation*}
\begin{aligned}
    V_{app}(\theta) &= E\left\{(H_0+ G_0 F_f(\theta))v_t)^2\right\}  \\
    &= \frac{1}{2\pi} \int_{-\pi}^\pi \left| H_0- G_0F_f(\theta) \right|^2 \Phi_v d\omega \leq \frac{1}{\gamma},
\end{aligned}
\end{equation*}
where $\gamma$ is some positive constant. The requirement is hence that the output variance should be less than $1/\gamma$. The overall objective of the input design is thus to design our input signal to be used during the experiment such that $V_{app}(\hat \theta_N)\leq 1/\gamma$.

We can approximate the application cost by a Taylor series expansion
\begin{equation*}
    V_{app}(\theta) \approx \frac{1}{2}\left[\theta-\theta_0\right]^T V_{app}'' \left[\theta-\theta_0\right],
\end{equation*}
and hence the set of all parameters, $\theta$ that satisfies $V_{app}(\theta) \leq 1/\gamma$ can be approximated by the ellipsoidal set
\begin{equation}
    \mathcal{E}_{app} = \left\{\theta : \frac{1}{2}\left[\theta-\theta_0\right]^T V_{app}'' \left[\theta-\theta_0\right] \leq \frac{1}{\gamma}  \right\}.
    \label{eqn:Vapp_set}
\end{equation}

\subsection{Minimum Variance Input Design}
To satisfy the application constraint we need to insure that the identified parameters lies within the application set (\ref{eqn:Vapp_set}) with high probability, \ie that $\mathcal{E}_{SI} \subseteq \mathcal{E}_{app}$. This is equivalent to
\begin{equation*}
    \frac{N}{\kappa} P^{-1} \succeq \gamma V_{app}''(\theta_0),
\end{equation*}
where $A \succeq B$ means that $A-B$ is positive semidefinite.
If the above inequality holds then $\hat \theta \in \mathcal{E}_{app}$ with at least probability $\alpha$.

The objective here is to find the minimum variance input signal used during the system identification experiment such that the identified model will satisfy the requirements from the application.
More formally this can be stated as the following optimization problem
\begin{equation}
\begin{aligned}
&\underset{\Phi_u,\Phi_{ue}}{\text{minimize}}  && \frac{1}{2\pi} \int_{-\pi}^\pi \Phi_u(\omega) d\omega\\
&\text{subject to} && \frac{N}{\kappa} P^{-1} \succeq \gamma V_{app}''(\theta_0) \\
& && \Phi_{\chi_0} \text{ defines a spectrum.}
\end{aligned}
\label{eqn:optprob}
\end{equation}

Assume that we can express the constraint that $\Phi_{\chi_0}$ defines a spectrum as an \LMI-constraint. Then since in many cases, as we will se in the following, $P^{-1}$ is an affine function of the spectral density $\Phi_{\chi_0}$ and since $\frac{1}{2\pi} \int_{-\pi}^\pi \Phi_u(\omega) d\omega$ is linear in $\Phi_{\chi_0}$, the optimization problem (\ref{eqn:optprob}) becomes a \SDP and can be solved efficiently.
We will come back to how to write the spectrum constraint as an \LMI later.

\section{First order \FIR subsystems}\label{sec:FIR}
To get some insight in the problem we start by looking at a low order example where we can solve the optimization problem analytically. We assume that the two subsystems are first order \FIR-systems, \ie
\begin{equation*}
    \begin{aligned}
    G_0(q) = 1 + b_0 q^{-1},\,
    H_0(q) = 1 + h_0 q^{-1},
    \end{aligned}
\end{equation*}
and the measurable disturbance is zero mean white noise with variance $\lambda_v$, \ie that $M=1$ and $\lambda_v = \lambda_s$. We will use system identification to identify the parameters $\theta=\begin{bmatrix}b & h\end{bmatrix}^T$ and use this to design a feed-forward controller.

The feed-forward controller will be on the form
\begin{equation}
    F_f(q,\theta) = \frac{c_0 + c_1 q^{-1}}{1+a_0{q^-1}},
    \label{eqn:feed-forward_form}
\end{equation}
where the constants $c_0$, $c_1$ and $a_0$ are functions of $h$ and $b$. Assuming that we know the true underlying system we design the optimal feed-forward controller, on the form (\ref{eqn:feed-forward_form}), that minimizes the output variance due to the disturbance $v_t$. If $G(q)$ is minimum phase then obvious we should invert $G$ in the feed-forward filter. In the non-minimum phase case it is known from minimum variance control that one should mirror the zero and then invert the system, see \cite{Astrom:1984}. The optimal parametrization is thus given by
\begin{equation*}
    \begin{bmatrix}c_0 & c_1 & a_0 \end{bmatrix} =
    \left\{
    \begin{matrix}
    \begin{bmatrix} 1 & h &  b\end{bmatrix} & \text{if } |b| < 1 \\
    \begin{bmatrix} \frac{hb^2+b-h}{b^3} & \frac{h}{b^2} &  \frac{1}{b}\end{bmatrix} & \text{if } |b| >1.
    \end{matrix}
    \right.
\end{equation*}

The objective of the system identification is to find an estimate $\hat \theta_N$ of $\theta$ such that when they are used to design a feed-forward controller $F_f(\hat \theta_N)$ the output variance due to the disturbance is less than $1/\gamma$. All parameters, $\theta$, that satisfies this is given by
\begin{equation}
    V_{app}(\theta) = E \left\{((H(\theta_0)+G(\theta_0)F_f(\theta))v_t)^2\right\} \leq \frac{1}{\gamma}.
    \label{eqn:app_cost}
\end{equation}

\subsection{Minimum phase system}
First we look at the case when $G_0$ is minimum phase, \ie $|b| <1$. The application cost (\ref{eqn:app_cost}) can now be calculated, using for example residue calculus, to
\begin{equation*}
    V_{app}(\theta) = \frac{c^2+d^2-2cd b}{1- b}\lambda_v \leq \frac{1}{\gamma},
\end{equation*}
where
$c = h_0 - h + b - b_0$, $d = h_0 b - b_0 h$.
The Hessian is then given by
\begin{equation*}
V_{app}^{\prime\prime}(\theta_0) = 2\lambda_v\begin{bmatrix} p & -1 \\ -1 & 1\end{bmatrix}, \quad p = \frac{h_0^2-2h_0b_0+1}{1-b_0^2}.
\end{equation*}

Using (\ref{eqn:pinv}) the asymptotic covariance matrix for the identified parameters can be calculated as
\begin{equation*}
    P^{-1} = \frac{1}{2\pi} \int_{-\pi}^\pi I \begin{bmatrix}\Phi_u & \Phi_{uv} \\ \Phi_{uv}^* & \lambda_v \end{bmatrix} I d\omega =
    \begin{bmatrix}
        r_u(0) & r_{uv}(0) \\
        r_{uv}(0) & \lambda_v
    \end{bmatrix},
\end{equation*}
where $r_u(k) = E[u_tu_{t-k}]$ and $r_{uv}(k) = E[u_tv_{t-k}]$ are the autocovariance of $u_t$ and the cross covariance between $u_t$ and $v_t$, respectively.

Since the input power can be written as
\begin{equation*}
    \frac{1}{2\pi} \int_{-\pi}^\pi \Phi_u(\omega) d\omega = r_u(0),
\end{equation*}
the optimization problem (\ref{eqn:optprob}) can be reformulated as
\begin{equation*}
\begin{aligned}
&\underset{r_u,r_{uv}}{\text{minimize}}  &&r_u(0)\\
&\text{subject to} && \frac{N}{\lambda_e\kappa} \begin{bmatrix} r_u(0) & r_{uv}(0) \\ r_{uv}(0) &  \lambda_v \end{bmatrix} \succeq \gamma \lambda_v\begin{bmatrix} p & -1 \\ -1 & 1 \end{bmatrix}.
\end{aligned}
\end{equation*}
Note that condition that $r_u$ and $r_{uv}$ corresponds to a realizable experiment (or that $\Phi_{\chi_0}$ defines a spectrum) is that the matrix $P^{-1}$ is positive semidefinite. However, since $V_{app}''$ is positive semidefinite this is already taken care of in the optimization problem.

The optimization problem above can be solved analytically, see \cite{Wahlberg:2012} for details. The optimization problem above is feasible if
\begin{equation*}
    \frac{N}{\lambda_e\kappa} - \gamma > 0
\end{equation*}
and the optimal solution is given by
\begin{equation}
\begin{aligned}
    r_u(0) &= \frac{ \gamma \lambda_e\lambda_v \kappa p }{N}, \\
    r_{uv}(0) &= -\frac{\gamma \lambda_e \lambda_v \kappa}{N}.
\end{aligned}
\label{eqn:opt_cor_mp}
\end{equation}

We are now ready to make the following observations:
\begin{itemize}
  \item To guarantee that the identified parameters satisfy the application requirement with high probability $\alpha$, we require $\frac{N}{\lambda_e\kappa} - \gamma > 0$. Hence the highest possible achievable accuracy for a given noise variance, experimental length and probability is
      \[ \gamma < \frac{N}{\lambda_e\kappa}.  \]
      Or conversely, to achieve a certain accuracy the experimental length need to satisfy
      \[ N > \gamma \lambda_e \kappa.  \]
  \item As expected the required input power during the identification increases with higher probability (larger $\kappa$), higher noise variance (larger $\lambda_e$) and tighter application requirements (larger $\gamma$), while increasing the experimental length, $N$, reduces the required power.
  \item Not as obvious is that the required input power is proportional to the disturbance variance. One might think that a higher variance of $v_t$ could help us to identify the parameter corresponding to the disturbance filter, $H$, and hence require less accuracy in the identification of $G$, consequently requiring less power in $u_t$. While it is true that a higher power of the disturbance makes the identification of $H$ more accurate, the application cost is also proportional to $\lambda_v$. Hence  higher accuracy is needed to be able to satisfy the application requirements.
  \item The required input power is proportional to $p$. The constant $p$ captures the effect of the underlying system,
      \begin{equation*}
        p=\frac{h_0^2-2h_0b_0+1}{1-b_0^2} = \frac{(h_0-b_0)^2}{1-b_0^2}+1 \geq 1.
      \end{equation*}
    Hence a lower bound on the required input power is given by
  \begin{equation*}
    r_u(0) \geq \frac{ \gamma \lambda_e\lambda_v \kappa  }{N}.
  \end{equation*}
  \item If the two systems are equal, \ie $b_0 = h_0$, then $p= 1$. Thus the least amount of power in the input is needed if the two systems are equal. In system identification of structured systems it is recognized that it could be hard to identify subsystems which are equal, see \cite{Wahlberg:2009,Hagg:2011}. Here we see that in this particular scenario this is not the case. If $h_0$ and $b_0$ are very different from each other $p$ will be large and the required power is large. Furthermore we can se that if $b_0$ is close to $\pm 1$ then high power is required. This is expected since we invert $G_0$ in out feed-forward filter and in this case we are close to the stability margin and a high accuracy is needed.
  \item If it is known beforehand that the two subsystems are equal this should of course be taken into account. In this case the feed-forward filter becomes $F_f = -1$ and is independent of the identified parameters. Hence it is not necessary to perform any system identification.
  \item The optimal input signal should always be negatively correlated with the disturbance and be
    \[ r_{uv}(0) = -\frac{\gamma \lambda_e \lambda_v \kappa}{N},\]
    independent of the underlying system.
\end{itemize}

\subsection{Non-minimum phase system}
Now we look at the case when $G_0$ is a non-minimum phase system, \ie $|b| >1$.
The optimal feed-forward controller on the form (\ref{eqn:feed-forward_form}) will now be
\begin{equation*}
    F_f(q,\theta) = \frac{(h b^2+b -h)q+h b}{b^2(b q +1)}.
\end{equation*}
Even if we know the parameters of the true system the variance will not be zero in this case. The smallest error we can make is given by
\begin{equation*}
    V_{min} = \frac{(b_0^2-1)(b_0-h_0)^2}{b_0^4}.
\end{equation*}
The application cost is now given by
\begin{equation*}
\begin{aligned}
V_{app}(\theta) &=  E\left\{\left((H+G_1F_f)v_t\right)^2\right\} -V_{min}\leq \frac{1}{\gamma}.
\end{aligned}
\end{equation*}
Note that $1/\gamma$ does no longer correspond to the highest acceptable variance but is now a bound on how much higher the variance can be using the identified parameters compared to if we knew the true system.
Redoing the same calculations as in the minimum phase case, the optimal input signal is if
\begin{equation*}
\frac{N}{\lambda_e \kappa} -\gamma \frac{b_0^4-b_0^2+1}{b_0^4} \geq 0,
\end{equation*}
given by
\begin{equation*}
\begin{aligned}
    r_u(0)      &= \frac{\gamma \lambda_e \lambda_v\kappa}{N}p_2, \\
    r_{uv}(0)   &= -\frac{\gamma \lambda_e \lambda_v\kappa}{N}\frac{h_0(b_0^4-3b_0^2+4)+2b_0^3-3b_0}{b_0^5},
\end{aligned}
\end{equation*}
where
\begin{equation}
    \begin{aligned}
    p_2 =& \frac{h_0^2(b_0^6+16b_0^2-6b_0^4-10)}{b_0^6(b_0^2-1)} \\
    &+\frac{h_0(4b_0^5-18b_0^3+12b_0)+4b_0^4-3b_0^2}{b_0^6(b_0^2-1)}.
    \end{aligned}
    \label{eqn:p2}
\end{equation}

Many of the observations for the minimum phase setting also hold in this case, the differences are
\begin{itemize}
    \item To achieve a certain accuracy the experimental length must now satisfy
      \[ N > \gamma \lambda_e \kappa\frac{b_0^4-b_0^2+1}{b_0^4}.\]
    \item The required input power is proportional to $p_2$ defined in (\ref{eqn:p2}). We see that, for a fixed $h_0$, if $|b_0| \rightarrow \infty$ then
        \begin{equation*}
            p_2 \rightarrow 0 \quad \Rightarrow \quad r_u(0) \rightarrow 0.
        \end{equation*}
        Thus a larger $b_0$ requires less input power to achieve a certain accuracy.
    \item If $b_0=h_0$ then
    \begin{equation*}
    \begin{aligned}
        r_u(0)      &= \frac{\gamma \lambda_e \lambda_v\kappa}{N} \frac{b_0^4-b_0^2+1}{b_0^4}, \\
        r_{uv}(0)   &= -\frac{\gamma \lambda_e \lambda_v\kappa}{N}\frac{b_0^4-b_0^2+1}{b_0^4}. \\
    \end{aligned}
    \end{equation*}
    \item Again if $|b_1| \rightarrow 1$ then $r_u(0) \rightarrow \infty$. This is at the stability margin of our feed-forward controller and it is required that we identify $G$ correctly as minimum phase or as non-minimum phase. Thus high accuracy and consequently high input power is needed.
\end{itemize}

\subsection{Signal generation}
We will now show how we can realize a input signal from the optimal correlations. Here we will only show how to do this for the minimum-phase case as the calculations for the non-minimum phase case is analogous. From the optimal correlations (\ref{eqn:opt_cor_mp}) we can generate an input signal by using
\begin{equation*}
    u_t = Kv_t + r_t,
\end{equation*}
where $r_t$ is a white zero mean Gaussian process independent of $v_t$ with variance $\lambda_r$ . The constant $K$ can in this case be seen as a feed-forward filter that should be used during the identification. Since
\begin{equation*}
\begin{aligned}
    r_u(0) &= K^2\lambda_v + \lambda_r, \\
    r_{uv}(0) &= K\lambda_v,
\end{aligned}
\end{equation*}
we obtain
\begin{equation*}
\begin{aligned}
    K &= \frac{r_{uv}(0)}{\lambda_v} = -\frac{\gamma \lambda_e \kappa}{N}, \\
    \lambda_r &= r_u(0)-K^2 \lambda_v = \lambda_v\left(\frac{\gamma \lambda_e\kappa p}{N} - \frac{\gamma^2\lambda_e^2\kappa^2}{N^2} \right) \geq 0,
\end{aligned}
\end{equation*}
where the last inequality is due to that $N > \gamma \lambda_e \kappa$ to have a feasible optimization problem.

\subsection{White noise input}
Let us compare the results with what we get if we use a white input signal during the experiment, uncorrelated with the measured disturbance $v_t$, \ie $r_{uv}(0) = 0$.
Now the optimization problem becomes
\begin{equation}
\begin{aligned}
&\underset{r_u,r_{uv}}{\text{minimize}}  &&r_u(0)\\
&\text{subject to} && \frac{N}{\lambda_e\kappa} \begin{bmatrix} r_u(0) & 0 \\ 0 &  \lambda_v \end{bmatrix} \succeq \gamma \lambda_v\begin{bmatrix} p & -1 \\ -1 & 1 \end{bmatrix}
\end{aligned}
\end{equation}
with optimal input
\begin{equation*}
    r_u(0) = \frac{ \gamma \lambda_e\lambda_v \kappa p }{N} + \frac{ \gamma^2 \lambda_e^2 \kappa^2 \lambda_v p }{N(N-\gamma\lambda_e\kappa)}.
\end{equation*}
Comparing to the case with correlated input the required power is
\begin{equation*}
     \frac{ \gamma^2 \lambda_e^2 \kappa^2 \lambda_v p }{N(N-\gamma\lambda_e\kappa)}
\end{equation*}
larger, see (\ref{eqn:opt_cor_mp}). The difference decays as $1/N^2$ so for large $N$ one could instead use an input signal uncorrelated with $v_t$.

\section{The general case} \label{sec:general}
We will now study more general cases, when the subsystems are not necessary \FIR-filters. Mainly we will focus the discussion on how to guarantee that the spectrum we optimize over, $\Phi_{\chi_0}$ actually defines a spectrum, and how to formulate this as an \LMI. Two common ways to do this are to use a finite dimensional parametrization or a partial correlation parametrization of the spectrum.

In the finite dimensional parametrization the spectrum is written as an infinite series and the optimization parameters are given by the truncated vector of coefficients in the series. Using the Kalman-Yakubovich-Popov lemma the condition that the parameters represent a spectrum can be written as an \LMI, see \cite{Jansson:2004} for details.

Often the optimization problem (\ref{eqn:optprob}) can be written in terms of only a finite number of parameters. The idea of the partial correlation approach is to find conditions on these parameters that guarantee the existence of an infinite extension such that the complete sequence defines a spectrum. This can also be expressed as an \LMI see \cite{Jansson:2004}.

The two above approaches are however not directly applicable to our problem. The problem is that we do not have control over the second input $v_t$. For example, the partial correlation approach only guarantees that there exist an extension, not that the extension exactly corresponds to the given properties of $v_t$.

\begin{figure}
\centering
\subfloat[Feedback]{\label{fig:feedback_a}\pgfplotsset{width=0.2\columnwidth,height=0.1\columnwidth,compat=newest,plot coordinates/math parser=false}\tikzstyle{block} = [draw, fill=kthgreen!20, rectangle,
    minimum height=3em, minimum width=3em,align=center]
\tikzstyle{sum} = [draw, fill=kthgreen!20, circle, node distance=1cm]
\tikzstyle{input} = [coordinate]
\tikzstyle{output} = [coordinate]
\tikzstyle{pinstyle} = [pin edge={to-,thin,black}]

\begin{tikzpicture}[auto, node distance=1cm,>=latex',transform shape,scale=0.6]

    \node [input] (inputr) {};
    \node [sum,right of=inputr,node distance=1.5cm] (sum) {$+$};
    \node [block,right of=sum,node distance=1.5cm] (G0) {$G_0$};
    \node [sum,right of=G0,node distance=1.5cm] (sum2) {$+$};
    \node [block,above of=sum2,node distance=1.5cm] (H0) {$H_0$};
    \node [input,above of=H0] (inputv) {};
    
    \node [block,below of=G0,node distance=1.5cm] (K) {$K$};
    \node [output,right of=sum2] (out1){};
    
    \draw [draw,->] (inputr) -- node[]{$r$} (sum);
    \draw [draw,->] (sum) -- node[]{$u$} (G0);
    \draw [draw,->] (G0) --  (sum2);
    \draw [draw,->] (inputv) -- node[]{$s$} (H0);
    \draw [draw,->] (H0) --  node[]{$v$}(sum2);
    \draw [draw,->] (sum2) -- node[name=y]{$y$}(out1);
    \draw [draw,->] (y) |- (K);
    \draw [draw,->] (K) -| node[very near end]{$-$}(sum.south);
    
\end{tikzpicture}}
\subfloat[Feed-forward]{\label{fig:feedback_b}\pgfplotsset{width=0.2\columnwidth,height=0.1\columnwidth,compat=newest,plot coordinates/math parser=false}\tikzstyle{block} = [draw, fill=kthgreen!20, rectangle,
    minimum height=3em, minimum width=3em,align=center]
\tikzstyle{sum} = [draw, fill=kthgreen!20, circle, node distance=1cm]
\tikzstyle{input} = [coordinate]
\tikzstyle{output} = [coordinate]
\tikzstyle{pinstyle} = [pin edge={to-,thin,black}]

\begin{tikzpicture}[auto, node distance=1cm,>=latex',transform shape,scale=0.6]

    \node [input] (inputr) {};
    \node [sum,right of=inputr,node distance=1.5cm] (sum) {$+$};
    \node [output,right of=sum,node distance=1.5cm] (G0) {};
    \node [sum,right of=G0,node distance=1.5cm] (sum2) {$+$};
    
    \node [block,above of=sum2,node distance=1.5cm] (H0) {$-M$};
    \node [input,above of=H0] (inputv) {};
    
    \node [block,below of=G0,node distance=1.5cm] (K) {$K$};
    \node [output,right of=sum2] (out1){};
    
    \draw [draw,->] (inputr) -- node[]{$r$} (sum);
    \draw [draw,->] (sum) -- node[]{$u$} (G0);
    \draw [draw,->] (inputv) -- node[]{$s$} (H0);
    \draw [draw,->] (H0) -- node[]{$v$}(sum2);
    \draw [draw,->] (sum2) -- node[name=y]{$y$}(out1);
    \draw [draw,->] (y) |- (K);
    \draw [draw,->] (K) -| node[very near end]{$-$}(sum.south);
    
\end{tikzpicture}}
\caption{The relation between the input $u_t$ in the closed loop case and the feed-forward case. }
\end{figure}
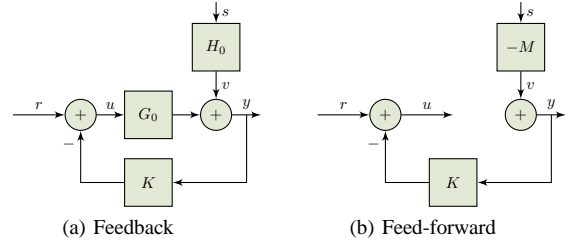

The two approaches have however been extended to closed loop optimal input design, see for example \cite{Hjalmarsson:2008} for the finite dimensional parametrization and \cite{Hildebrand:2010} for the partial correlation approach. We will show that our problem is just a special case of this. In closed loop input design the objective is to find the controller $K$ and the spectrum of the reference signal $r_t$, with $u_t = -Ky_t + r_t$, to achieve some properties on the identified models. See \fig{fig:feedback_a}. This is the same as designing the spectrum of $u$, $\Phi_u$ and the cross spectrum between $u_t$ and $s_t$, $\Phi_{us}$.
This is exactly what we want to do in the feed-forward case, we want to find conditions such that $\Phi_u$ and $\Phi_{us}$ corresponds to realizable experiment. We see that the feed-forward case in \fig{fig:feedback_b} corresponds to the closed loop case when $G_0 = 0$ and $H_0 = -M$. Hence we can use the existing results from closed loop optimal input design to guarantee that the spectra in the optimization can be realized by a reference signal $\Phi_r$ and a controller $K$ with $u_t = -Ky_t +r_t = Kv_t + r_t$.
Here we will show how the partial correlation approach from closed loop optimal input design \cite{Hildebrand:2010} can be used in this setting.

First we define the generalized moments of the spectrum $\Phi_{\chi_0}$ as
\begin{equation}
    m_k = \frac{1}{2\pi} \int_{-\pi}^\pi \frac{1}{|d(e^{j\omega})|^2} \Phi_{\chi_0} e^{jk\omega} d\omega,
    \label{eqn:gen_moments}
\end{equation}
where $d(z) = \sum_{l=0}^m d_l z^l$ is such that the coefficients are real and obey $d_m \neq 0$ and has all its roots outside the closed unit disc. Note that $m_k$ is real and that $m_{-k} = m_k^T$

Many common cost functions in input design, for example minimum input variance, can be written as linear functions of the generalized moments, see \cite{Hildebrand:2010}. If the model structures of $G(q,\theta)$ and $H(q,\theta)$ are rational then by choosing $d(z)$ as the least common denominator of $\frac{\partial G(z,\theta)^*}{\partial\theta}$, $\frac{\partial H(z,\theta)^*}{\partial\theta}$ and $M^*(z)$ evaluated at $\theta = \theta_0$ we see that $P^{-1}$ is affine in the generalized moments $m_0,m_1,\ldots,m_n$ \cf (\ref{eqn:pinv}), see \cite{Jansson:2004} for details.

The input design problem (\ref{eqn:optprob}) can hence be written as a \SDP in a finite number of generalized moments (\ref{eqn:gen_moments}). Now we want to classify the set of all sequences $m_0,\ldots,m_n$ that corresponds to a realizable experiment design, \ie that corresponds to a $K$ and $\Phi_r$ that are realizable. In \cite{Hildebrand:2010} a semi-definite description of this set is given. We will shortly summarize the result here.

The sequence $(m_0,m_1,\ldots,m_n)$ of $2\times2$ matrices satisfying $m_{-k}=m_k^T$ for $k=0,\ldots,n$ defines a valid experiment if the following conditions hold\footnote{$m_{k,pq}$ denotes the $(p,q)$ element of the matrix $m_k$.}
\begin{enumerate}
  \item $m_{k,22} = \frac{1}{2\pi} \int_{-\pi}^\pi \frac{\lambda_s}{|d(e^{j\omega})|^2} e^{jk\omega}$ for $k=-n,\ldots,n$.
  \item $\sum_{l=0}^m d_l m_{k-l,21} = 0$ for $k=1,\ldots,n$.
  \item The matrix
  \begin{equation*}
    T_n =
    \begin{bmatrix}
        m_0 & m_1^T & \cdots & m_n^T \\
        m_1 & m_0^T & & m_{n-1}^T \\
        \vdots & & \ddots & \vdots \\
        m_n & m_{n-1} & \cdots & m_0
    \end{bmatrix}
  \end{equation*}
   is positive semi-definite.
\end{enumerate}
A proof and more details can be found in \cite{Hildebrand:2010}.

\begin{remark}
The above formulation is not only useful for feed-forward cases. It could as well be applied, with minor modifications, to more general \MIMO system identification problems where some of the inputs are known but not controllable. This is a common case in many industrial applications.
\end{remark}

\subsection{Generating the Input Signal}\label{sec:generate_input}
Solving the optimization problem gives the optimal $m_0,m_1,\ldots,m_n$. The problem is then to find the
feed-forward controller $K$ and the spectrum $\Phi_r$ from theses matrices.

Define the matrix $T_n^r(a) = \text{diag}(a,0,a,0,\ldots,a,0)$ and let $a_{max}$ be the largest possible value of $a$ such that $T_n - T_n^r(a)$ is positive semi-definite. Furthermore let $\mathbf{v}$ be a non-zero vector such that $(T_n - T_n^r(a_{max}))\mathbf{v} = 0$ with $\mathbf{v} = (p_n,q_n,p_{n-1},q_{n-1},\ldots,p_0,q_0)^T$. Then one possible realization of the reference spectrum and the feed-forward controller is
\begin{equation*}
    \begin{aligned}
        \Phi_r(\omega) &= |d(e^{j\omega})|^2 a_{max}, \\
        K &= -\frac{q(z)}{p(z)M(z)},
    \end{aligned}
\end{equation*}
where $p(z) = \sum_{l=0}^n p_l z^{-l}$ and $q(z) = \sum_{l=0}^n q_l z^{-l}$. The spectrum of $r$ can thus be realizable by filtering white noise with variance $a_{max}$ through the filter $d^*(z)$.
Other realizations are possible, see \cite{Hildebrand:2010} for details and a proof.

\section{Example - House Heating}\label{sec:example}
In this section we will show how the framework for input design for feed-forward control can be applied in a simulation example. We will consider the problem of controlling the temperature in a room using electrical radiators. The control signal $u_t$, is the temperature of the radiator and $y_t$ is the temperature in the room that we would like to control. The temperature in the room is also influenced by the outside temperature, $v_t$, due to heat transfer in the walls. To simplify the problem we consider $v_t$ as white noise with variance $\lambda_v$. A thermometer is fitted outdoors so we assume that we can measure the temperature outside perfectly. The model of the system we will use here is a simplified version of the one given in \cite{Glad:2000}. The model is given by
\begin{equation*}
    y_{t+1} = -b y_t + k_1u_t+k_2v_t
\end{equation*}
where $b$, $k_1$ and $k_2$ are some constants that depend on the heat transfer coefficients between radiator and air, the coefficient through the walls and the sampling time.
This can be rewritten as
\begin{equation*}
    y_t = \underbrace{\frac{k_1}{1+bq^{-1}}}_G u_t + \underbrace{\frac{k_2}{1+bq^{-1}}}_Hv_t + e_t
\end{equation*}
where we have added $e_t$ as zero mean white Gaussian measurement noise with variance $\lambda_e$.

The objective is to identify the parameters $\theta=\begin{pmatrix}k_1,k_2,b \end{pmatrix}^T$ and design a feed-forward filter such that we keep the variance in the indoor temperature due to changing outdoor temperature less than $1/\gamma$.

The feed-forward filter based on the identified parameters will be
\begin{equation*}
    F_f(q,\theta) = -\frac{k_2}{1+ b q^{-1}}\left(\frac{ k_1}{1+ b q^{-1}}\right)^{-1} = -\frac{ k_2}{ k_1}.
\end{equation*}
Since both $G$ and $H$ have the same dynamics we see that the accuracy of the the parameter $b$ is not important. The application cost is hence
\begin{equation*}
\begin{aligned}
V_{app}(\theta) &= E\left\{((H_0+G_0F_f(\theta))v_t)^2\right\} \leq \frac{1}{\gamma}
\end{aligned}
\end{equation*}
which can be calculated analytically using for example residue calculus. From this we can then derive $V_{app}''(\theta_0)$.

We will now use the framework outlined in Section~\ref{sec:general} to formulate an optimal experiment for this case.
First we look at the gradient $\mathcal{F}_0$ defined in (\ref{eqn:predict_grad})
\begin{equation*}
    \mathcal{F}_0 =  \begin{bmatrix} \frac{\partial G(q,\theta)}{\partial\theta} & \frac{\partial H(q,\theta)}{\partial\theta}  \end{bmatrix} =
    \begin{bmatrix}
    \frac{1}{1+bq^{-1}} & 0 \\
    0 & \frac{1}{1+bq^{-1}} \\
    -\frac{k_1 q^{-1}}{(1+bq^{-1})^2} & -\frac{k_2 q^{-1}}{(1+bq^{-1})^2}
    \end{bmatrix}
\end{equation*}
and
\begin{equation*}
    P^{-1} = \frac{1}{2\pi \lambda_e}\int_{-\pi}^{\pi} \mathcal{F}_0 \Phi_{\chi_0}\mathcal{F}_0^* d\omega.
\end{equation*}
By defining $d(e^{j\omega}) = (1+be^{j\omega})^2$ and $m_k$ as in (\ref{eqn:gen_moments}) we can write
$P^{-1}$ as linear combinations of $m_k$, $k=-2,\ldots,2$.

The input signal energy can be written in terms of $m_k$ as
\begin{equation}
\begin{aligned}
    \frac{1}{2\pi}\int_{-\pi}^\pi \Phi_u(\omega) d\omega &=
     \frac{1}{2\pi}\int_{-\pi}^\pi \frac{|d(e^{j\omega})|^2}{|d(e^{j\omega})|^2} \Phi_u(\omega) d\omega \\
     &= \sum_{k=-2}^{2} \delta_k m_{k,11},
    \end{aligned}
    \label{eqn:input_energy}
\end{equation}
where $\delta_k$ are the coefficients of $|d(e^{j\omega})|^2$, \ie $|d(e^{j\omega})|^2 = \sum_{k=-2}^2 \delta_k e^{j\omega k}$.
In the same way we can express the output variance during the experiment as
\begin{equation}
    \frac{1}{2\pi}\int_{-\pi}^\pi \Phi_y(\omega) d\omega = \sum_{k=-1}^1 \beta_k \begin{bmatrix}k_1 & k_2 \end{bmatrix} m_k \begin{bmatrix}k_1 \\ k_2 \end{bmatrix},
    \label{eqn:output_energy}
\end{equation}
where $\beta_k$ are the coefficients of $|1+be^{-j\omega}|^2$.

We can now formulate the optimal input design problem where we want to find either a minimum variance input signal or a input signal that gives the lowest output variance while satisfying the application constraints with high probability. The optimization becomes
\begin{equation*}
\begin{aligned}
&\underset{m_k,k=-2,\ldots,2}{\text{minimize}}  && (\ref{eqn:input_energy}) \text{ or } (\ref{eqn:output_energy})\\
&\text{subject to} && \frac{N}{\kappa} P^{-1} \succeq \gamma V_{app}''(\theta_0) \\
& && m_{k,22} = \frac{1}{2\pi} \int_{-\pi}^\pi \frac{\lambda_v}{|d(e^{j\omega})|^2} e^{jk\omega}, \, \forall k \\
& && \sum_{l=0}^m d_l m_{k-l,21} = 0, \,  k=1,\ldots,2 \\
& && \begin{bmatrix} m_0 & m_1^T & m_2^T \\ m_1 & m_0 & m_1^T  \\ m_2 & m_1 & m_0 \end{bmatrix} \succeq 0.
\end{aligned}
\end{equation*}
The above problems are difficult, if not impossible, to solve analytically. As the problems are \textsc{sdp}s  they can be solved efficiently using numerical methods. The following system parameters will be used; $b = -0.5$, $\lambda_v = 3$, $\lambda_e = 1$, $k_1 = 1$ and $k_2 = 0.3$. The experimental conditions are $N=1000$, $\gamma = 100$ and $\kappa =5.99$ corresponding to at least $95\%$ probability that the identified model satisfies the application requirement.

\begin{figure}
\centering
{\pgfplotsset{width=\columnwidth,height=0.4\columnwidth,compat=newest,plot coordinates/math parser=false}
%
%
%
%
\begin{tikzpicture}

\begin{semilogxaxis}[%
xmin=0.01, xmax=3.14159265358979,
xminorticks=true,
ymin=0, ymax=0.7,
ylabel={Magnitude},
xlabel={Frequency (rad/s)},
name=plot1]
\addplot [
color=kthblue,
solid,
forget plot,
line width=1
]
coordinates{
 (0.00950705590450297,0.55709079596887)(0.0112097729549602,0.55665731126838)(0.0132174472269843,0.556056329306443)(0.0155846966660294,0.555224030677811)(0.0183759213107571,0.554073105399902)(0.0216670552693643,0.552484856955489)(0.0255476324754868,0.550299330633935)(0.0301232224217101,0.547303599372313)(0.0355183021338077,0.543219042710466)(0.0418796425165735,0.537689765210777)(0.0493803011953816,0.530276431434157)(0.0582243304770719,0.520462700455817)(0.0686523285082827,0.507684318198734)(0.0809479846481901,0.49139151637878)(0.0954457971198032,0.471149783308972)(0.112540172895315,0.446768684562432)(0.132696157374116,0.41842507693395)(0.156462085749908,0.386728646001787)(0.184484500242106,0.352684297763706)(0.217525739008624,0.317545928344299)(0.25648467523912,0.28260859226526)(0.302421170627118,0.249013271248633)(0.356584908467572,0.217622193948462)(0.42044939077233,0.188980987961148)(0.495752024280914,0.163348371747004)(0.584541385889906,0.140760734837129)(0.689232953337324,0.121103824855091)(0.812674817272153,0.104175442858035)(0.958225162378583,0.0897333827508261)(1.12984362539681,0.0775291612328804)(1.33219901539745,0.0673309756286612)(1.5707963267949,0.0589403778318693)(1.80437115662468,0.0531690426759736)(2.07267817941887,0.0485190191489999)(2.38088201513671,0.0449904192635814)(2.73491525422956,0.0426678646521342)(3.14159265358979,0.0417937928200997) 
};\label{lbl:Kinput}

\addplot [
color=kthgreen,
dashdotted,
forget plot,
line width=1
]
coordinates{
 (0.00138629436111989,0.599998846916275)(0.0138629436111989,0.599884726351875)(0.0159321724467921,0.599847760711847)(0.0183102612254201,0.599798947321023)(0.0210433114041914,0.599734493569662)(0.0241843056962509,0.599649396265779)(0.0277941351898266,0.599537057807965)(0.0319427797784625,0.599388782582068)(0.0367106647861749,0.599193117632898)(0.0421902200869685,0.598934991941935)(0.0484876719436905,0.598594597112942)(0.0557251023026812,0.598145939272549)(0.064042815465558,0.597554978608496)(0.0736020580182617,0.596777261621242)(0.0845881447457123,0.595754946636387)(0.0972140511308297,0.594413134092358)(0.111724542081845,0.592655454328487)(0.128400916927116,0.590358960545496)(0.147566462663561,0.58736855786582)(0.169592721175011,0.583491516276143)(0.194906691916301,0.578493113426332)(0.223999109693827,0.572095157458198)(0.257433958015019,0.563979996603077)(0.295859402431833,0.553803407205429)(0.34002035583128,0.541219964520665)(0.390772919262784,0.525923337661492)(0.449100978251228,0.507700548780083)(0.516135276330593,0.486493379588911)(0.593175335556352,0.462453250589753)(0.681714648946065,0.435971644811134)(0.783469633227032,0.407670841924301)(0.900412903166855,0.378350971906265)(1.03481151254069,0.348905146952266)(1.1892709030718,0.320226502274079)(1.36678541333643,0.293132989528224)(1.5707963267949,0.268328157299975)(1.80437115662468,0.246477208109899)(2.07267817941887,0.228014867549289)(2.38088201513671,0.213505759369219)(2.73491525422956,0.203726503696384)(3.14159265358979,0.2) 
};\label{lbl:H0}

\addplot [
color=kthred,
dashed,
forget plot,
line width=1
]
coordinates{
 (0.0031415926535898,0.3)(0.0314159265358979,0.3)(0.0353698983929485,0.3)(0.0398215125343506,0.3)(0.0448334016373536,0.3)(0.0504760812548838,0.3)(0.0568289419450803,0.3)(0.0639813662690947,0.3)(0.0720339863729318,0.3)(0.081100099847073,0.3)(0.0913072637845395,0.3)(0.102799089465244,0.3)(0.11573726291722,0.3)(0.130303819783332,0.3)(0.146703706500051,0.3)(0.165167663823206,0.3)(0.185955473270923,0.3)(0.209359612160082,0.3)(0.235709368661392,0.3)(0.265375474770513,0.3)(0.298775322379497,0.3)(0.33637883583691,0.3)(0.37871508362151,0.3)(0.426379722153464,0.3)(0.480043376474956,0.3)(0.540461075713481,0.3)(0.608482876081537,0.3)(0.685065820874663,0.3)(0.771287405740876,0.3)(0.868360738673208,0.3)(0.977651608021218,0.3)(1.1006976986625,0.3)(1.23923022669916,0.3)(1.39519829707187,0.3)(1.5707963267949,0.3)(1.80437115662468,0.3)(2.07267817941887,0.3)(2.38088201513671,0.3)(2.73491525422956,0.3)(3.14159265358979,0.3) 
};\label{lbl:Koutput}

\end{semilogxaxis}


\end{tikzpicture}%
}
\caption{The magnitude of the optimal feed-forward filters to be used during the identification for the minimum input variance case (\ref{lbl:Kinput}) and the minimum variance output (\ref{lbl:Koutput}). For reference $H_0$ is also shown as (\ref{lbl:H0}).}
\label{fig:bodeK}
\end{figure}
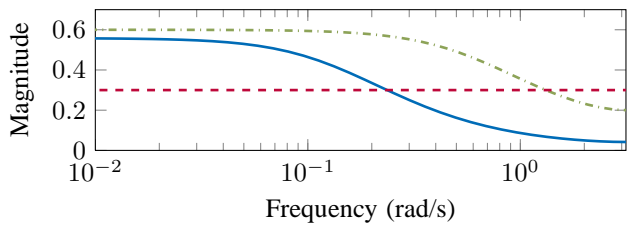

The two problems are solved with \CVX, a package for specifying and solving convex programs \cite{CVX:2012}, giving the optimal $m_k$, $k=-2,\ldots,2$. Using the algorithm outlined in Section~\ref{sec:generate_input} we generate a reference spectrum $\Phi_r$ and the feed-forward controller to be used during the identification, $K$. In both cases the optimal $T_n$ is singular and consequently $a_{max} = 0$ and $\Phi_r = 0$. Therefore it is enough to use $u_t = Kv_t$ during the identification. The magnitude of the optimal feed-forward controllers $K$ for both problems are shown in \fig{fig:bodeK} where also $H_0$ is shown for reference.
We see that when we minimize the input variance the optimal $K$ is a lowpass filter, while for the output variance case the optimal $K$ is a constant $K=-0.3$. The case when $K=-0.3$ corresponds to the optimal feed-forward controller $F_f = -k_2/k_1=-0.3$ that gives zero output variance in $y$ due to the measurable disturbance $v_t$.
But using this feed-forward filter during the identification gives that $y_t = e_t$, and hence we cannot say anything about the parameters $k_1$, $k_2$ and $b$ from this information. However, since we know that the subsystems have the same dynamics, this information is enough to identify the ratio $k_2/k_1$ which is all information we need to design our feed-forward controller!

We verify the results in a Monte-Carlo simulation. In each round system identification data is generated using the optimal identification feed-forward filter, $K$, and the parameters $k_1$, $k_2$ and $b$ are identified. The identified parameters are then used to design the feed-forward controller. Finally we calculate the output variance due to the measurable disturbance $v_t$ when the feed-forward filter $F_f(\hat \theta_N)$ is used in the application. In about $98\%$ of the simulations the output variance was less than $1/\gamma$, thus the results seem to be valid.

\section{Conclusions}\label{sec:conclusions}
In this paper we have considered optimal input design when the identified model will be used in a feed-forward control application. First, the first order \FIR-filter case was considered and some fundamental properties were observed.
Secondly a framework for optimal input design for feed-forward systems was presented and the relation to closed loop input design was discussed. The framework was then successfully applied to a numerical example.

Interesting extensions of this work would be to see what happens if, on top of the feed-forward, feedback is added as this is the most common case in practice. It would also be interesting to see if some of the results for the low order \FIR-case can be extended to more general structures.

We noted that the framework presented here can be extended to more general \MIMO system identification problems were we only can control a few of the inputs. It would be interesting to see if this can be used in identification of complex interconnected systems, see \cite{Dankers:2012}. For example if one node locally want to identify the dynamics of the network, how should it excite the system when it only can measure, and not affect, the inputs from its neighbors?


\bibliographystyle{IEEEtran}
\bibliography{references}

\end{document}